\documentclass[serif,twocolumn]{wiley-article}
\usepackage{graphicx}        
\usepackage{amsmath} 
\usepackage{bm}              
\usepackage{siunitx}         
\usepackage{physics}         
\usepackage{booktabs}        
\usepackage{multirow}        
\usepackage{xcolor}          
\usepackage{float}           
\usepackage{hyperref}        

\usepackage[numbers,sort&compress,super]{natbib}
\papertype{Original Article}
\paperfield{Journal Section}

\title{Room-Temperature Electric-Field Control of Anomalous Hall Effect in Py/BTO/LSMO Heterostructures}

\author[1]{Kusampal Yadav}
\author[1]{Kousik Das}
\author[2]{Aditya Raj}
\author[2]{Mainak Ghosh}
\author[3]{Abhishek Kumar}
\author[3]{Kartick Biswas}
\author[1]{Kalyan Sarkar}
\author[3]{Pavan Nukala}
\author[2\authfn{2}]{Sayantika Bhowal}
\author[1\authfn{1}]{Devajyoti Mukherjee}
\affil[1]{School of Physical Sciences, Indian Association for the Cultivation of Science, 2A and 2B Raja S.~C.~Mullick Road, Kolkata 700032, India}
\affil[2]{Department of Physics, Indian Institute of Technology Bombay,
Powai, Mumbai 400076, India}
\affil[3]{Centre for Nanoscience and Engineering, Indian Institute of Science, Bengaluru 560012, India}

\corraddress{Sayantika Bhowal, Devajyoti Mukherjee} 
\corremail{sbhowal@iitb.ac.in, sspdm@iacs.res.in}

\runningauthor{Yadav et al.}
\begin{document}

\begin{frontmatter}
\maketitle

\begin{abstract}
We demonstrate room temperature electric field control of the anomalous Hall effect in epitaxial Ni$_{80}$Fe$_{20}$ (Py)/BaTiO$_{3}$ (BTO)/La$_{0.7}$Sr$_{0.3}$MnO$_{3}$ (LSMO) thin film heterostructures grown on MgO and LaAlO$_3$ substrates. Substrate induced strain states generate distinct magnetic anisotropies, enabling voltage driven tuning between anomalous and topological Hall contributions. Robust ferroelectric polarization in BTO, confirmed by piezoresponse force microscopy, couples strongly to interfacial orbital reconstruction and carrier redistribution. As a result, Hall resistivity exhibits giant low voltage tunability, with up to $\sim$93$\%$ modulation at operating voltages of only 0.5–2 V. Density functional theory calculations further reveal polarization controlled Rashba spin splitting, establishing a direct link between ferroelectric order and emergent quantum transport. These findings establish Py/BTO/LSMO heterostructures as promising candidates for low-power multifunctional spintronic devices, where substrate engineering enables control over emergent quantum transport phenomena.

\end{abstract}
\end{frontmatter}
\twocolumn
\section{Introduction}
Complex oxide–metal heterostructures have emerged as a powerful platform for manipulating interfacial strain, charge transfer and spin–orbit interactions, thereby enabling the emergent phenomena not observed in the corresponding bulk phases\cite{Zhang2010,Peng2019,Maccherozzi2025}. At these interfaces, the breaking of spatial inversion and time-reversal symmetry, together with lattice distortions arising from epitaxial mismatch and strong spin–orbit coupling (SOC), produce pronounced anomalous Hall effect (AHE) and stabilizes unconventional magnetic textures such as interfacial ferromagnetism, chiral spin configurations, and polar skyrmion phases that give rise to topological Hall signatures\cite{Begue2025,Maccherozzi2025,Jaiswal2022,Shao2023,Wang2018,Maity2024}. The strong coupling among electric polarization, charge transfer, and spin order further enables room-temperature electric-field control of magnetic and transport behavior, offering an attractive pathway toward low-power spintronic technologies. These possibilities shows that oxide–metal heterostructures are promising building blocks for next-generation non-volatile memories, magnetic memory devices, energy efficient nanoelectronics systems\cite{Ohuchi2018,Feng2020,chen2026,Yau2015}.\\
 We have recently reported that epitaxial thin film of Permalloy (Ni$_{80}$Fe$_{20}$, Py) a soft ferromagnet with high permeability and high Curie temperature, grown on half metallic ferromagnetic La$_{0.7}$Sr$_{0.3}$MnO$_{3}$ (LSMO) having near room temperature Curie temperature  exhibit a giant topological Hall resistivity at room temperature – nearly five times larger than single Py film\cite{Yadav2025}. This enormous enhancement is due to the combination of LSMO’s robust ferromagnetism and a strong Rashba-like spin–orbit interaction at the Py/LSMO interface. Remarkably, inserting a ferroelectric (FE) BaTiO$_3$ (BTO) layer between Py and LSMO does not suppress this effect; instead, the BTO spacer also shows the enhanced Hall signature. Theoretical modeling suggest that these emergent Hall contributions arise from broken inversion symmetry and spin reorientation at the oxide–metal boundary, which stabilize chiral spin textures and associated Berry-curvature fields in the Py layer\cite{Yadav2025}.\\
Expanding on these findings, it is envisaged that applying a voltage can modify the interface and enable electrical control of the Hall effect within the ferromagnetic layer. Electric-field control of magnetotransport in conventional metallic ferromagnets has largely focused on magnetization or anisotropy, however direct tuning of the anomalous Hall effect (AHE) remains challenging \cite{Begue2025_adv_Mater,Maccherozzi2025,Nan2014,Fackler2014,Bocirnea2020,Begue2025}. Ferroelectric oxides provide a promising pathway because of their switchable polarization which can modify interfacial symmetry, charge distribution, and orbital hybridization. Gate-responsive Hall signals have been reported in oxide heterostructures such as BaTiO$_3$/SrTiO$_3$, BaTiO$_3$/SrRuO$_3$, SrRuO$_3$/SrIrO$_3$, and BiFeO$_3$/SrRuO$_3$, where polarization and strong spin–orbit coupling generate changes in AHE or topological Hall signatures. However, these phenomena are mostly reported at low temperatures (<50–100 K)  \cite{Ohuchi2018,Wang2018, Shao2023,Ren2020}, limiting their direct relevance for room-temperature spintronic applications.
In contrast, control of room-temperature anomalous Hall resistivity in simple metal/ferroelectric systems has been relatively rarely reported. Although NiFe/PMN-PT and Ni/BaTiO$_3$ heterostructures have exhibited strain- and charge-mediated magnetoelectric coupling \cite{Nan2014, Bocirnea2020}, there have been no reports addressing the associated changes in Hall transport. Likewise, in Py/AlO$_X$/SrTiO$_3$ systems, the application of an external electric field has been shown to induce Rashba-driven spin–charge interconversion, arising from the ferroelectric-like behavior of STO below 100 K \cite{Noel2020}. Consequently, despite notable advancements, room-temperature ferroelectric control of AHE in simple metallic ferromagnets remains elusive. Importantly, Py/BaTiO$_3$ bilayer have been used to tailor Py anisotropy via strain \cite{Fackler2014,Begue2025,Begue2025_adv_Mater}, no studies have examined how BaTiO$_3$ or BiFeO$_3$ polarization can directly tune the Berry-curvature-driven AHE in Py. This leaves a clear gap in understanding how ferroelectric polarization—beyond strain alone—can act as an active gate to modulate Hall transport in metal/oxide heterostructures at room temperature.\\
In the present work we address these gaps by fabricating metal–oxide trilayers of Py/BTO/LSMO that operate robustly at room temperature. By combining a soft ferromagnetic metal (Py) with a strong perovskite ferroelectric (BTO) and a half-metallic manganite (LSMO), we demonstrate that the ferroelectric polarization can act as a nonvolatile gate to modulate Hall transport in the ferromagnetic layer. Application of an out-of-plane electric field switches the BTO polarization, thereby modifying interfacial charge accumulation, orbital hybridization, inversion symmetry, and Rashba-type spin–orbit coupling at the Py/BTO interface, which collectively alter the anomalous Hall resistivity. Modifying the Rashba-type spin-splitting in BTO by controlling ferroelectric polarization is demonstrated from the theoretical computations based on density functional theory. In these theoretical computations, the ferroelectric polarization of BTO is controlled by changing the atomic displacements from the inversion-symmetric positions. Furthermore, by employing lattice-mismatched MgO and LAO substrates, we systematically tune the epitaxial strain and consequently the interfacial magnetic anisotropy, ferroelectric polarization state, and topological transport response. Remarkably, the voltage-dependent transport measurements reveal giant electric-field control of Hall transport, achieving nearly $\sim$93$\%$ change of Hall resistivity at exceptionally low operating voltages of only $\sim$0.5 V and $\sim$2 V. Such large tunability at room temperature significantly exceeds most previously reported metallic ferromagnetic systems \cite{Yang2018, Gao2016, SONG2017315, Feng2020, CHEN20118885, Shimizu2013, Cai2025} and highlights the strong interplay among strain, polarization-driven charge redistribution, and interfacial spin–orbit coupling in these oxide heterostructures. The demonstrated coexistence of electrically tunable anomalous and topological Hall responses, together with substrate-controlled magnetic anisotropy, establishes Py/BTO/LSMO heterostructures as promising candidates for low-power multifunctional tangible spintronic applications.\\

\section{Results and Discussion}
\subsection{Crystallinity}

\begin{figure*}[h]
 \centering
    \includegraphics[width=2\columnwidth]{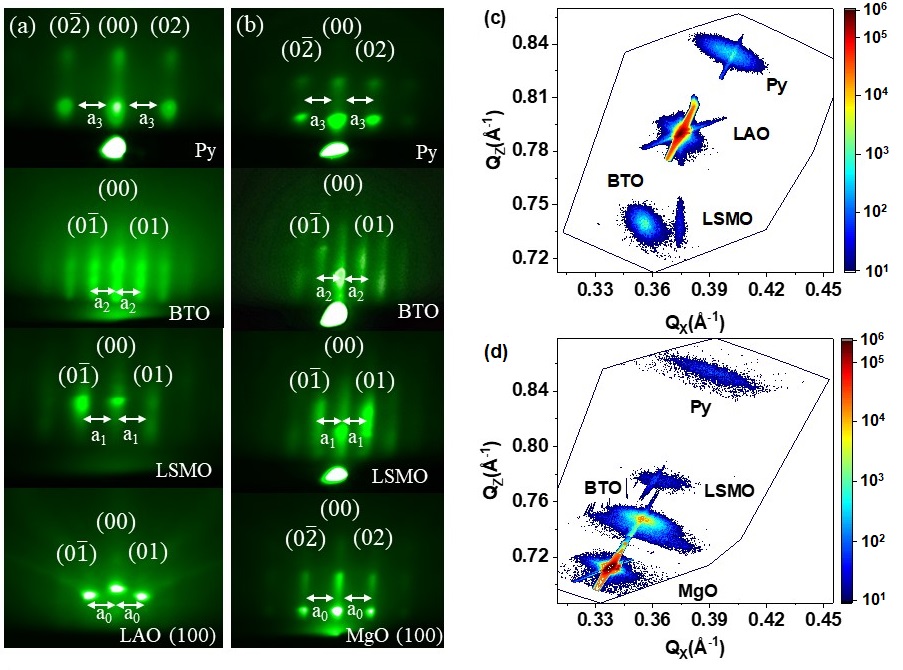}
    \caption{Structural characterization of epitaxial Py/BTO/LSMO heterostructure deposited on LAO and MgO substrates. (a–b) In-situ reflection high-energy electron diffraction (RHEED) patterns recorded from the Py, BTO, and LSMO layers for Py/BTO/LSMO/LAO and Py/BTO/LSMO/MgO heterostructures, respectively. (c–d) Reciprocal space maps (RSMs) measured around the (113) reflections for Py/BTO/LSMO/LAO and Py/BTO/LSMO/MgO heterostructures, respectively.}
    \label{fig_xrd}
\end{figure*} 
Figure \ref{fig_xrd} shows the structural characteristics of the epitaxial Py/BTO/LSMO heterostructures grown on LAO (100) (cubic, a = 3.79 $\si{\angstrom}$) and MgO (100) (cubic, a = 4.22 $\si{\angstrom}$) substrates. The Py layer (face-centered cubic, a = 3.55 $\si{\angstrom}$, JCPDS No. 01-088-9591) is deposited on tetragonal BTO (a $\sim$ 3.99 $\si{\angstrom}$, c $\sim$ 4.03 $\si{\angstrom}$, JCPDS No. 01-074-1957), which is grown on pseudocubic LSMO (a $\sim$ 3.88 $\si{\angstrom}$, JCPDS No. 01-089-4461), forming a cube-on-cube epitaxial multilayer structures on both substrates. The lattice mismatch between the layers and substrates provides a platform to tailor strain and interfacial coupling in the heterostructures. Fig.\ref{fig_xrd} (a-b) shows the in situ RHEED patterns of the Py, BTO, and LSMO layers along with those of the bare substrates LAO and MgO, respectively. The patterns exhibit sharp, streak-like features with well-defined Kikuchi lines, indicating smooth surfaces and high crystalline quality of all layers. The persistence of these streaks throughout the growth confirms a layer-by-layer growth mode and two-dimensional epitaxy of (100)-oriented films on both LAO and MgO substrates \cite{Wang2012,Yin2006, Yuan2015}. Quantitative analysis of the RHEED patterns (Fig. \ref{fig_xrd} (a,b)) allows estimation of the in-plane lattice parameters of the individual layers on both substrates. For the Py/BTO/LSMO/LAO heterostructure, the extracted values for Py, BTO, and LSMO are 3.61 $\pm$ 0.033 $\si{\angstrom}$, 3.94 $\pm$ 0.039 $\si{\angstrom}$, and 3.80 $\pm$ 0.037 $\si{\angstrom}$, respectively. In comparison, the corresponding lattice parameters for the Py/BTO/LSMO/MgO heterostructure are 3.62 $\pm$ 0.088 $\si{\angstrom}$, 3.97 $\pm$ 0.019 $\si{\angstrom}$, and 3.97 $\pm$ 0.086 $\si{\angstrom}$, indicating a clear substrate-dependent variation in the in-plane lattice constants. The single-crystalline nature of the Py/BTO/LSMO heterostructures is confirmed by high-resolution X-ray diffraction reciprocal space mapping (RSM) around the asymmetric (113) reflections (Fig. \ref{fig_xrd}(c,d)). The RSMs of both LAO- and MgO-based heterostructures exhibit well-defined diffraction spots corresponding to the fcc Py layer, pseudocubic LSMO, tetragonal BTO, and the respective substrates, confirming  the cube-on-cube type epitaxial growth. In case of Py/BTO/LSMO/LAO, the LSMO peak is closely aligned with the substrate reflection, consistent with the small lattice mismatch ($\sim$2.3 $\%$), indicating that LSMO layer in this case highly strained. In contrast, for the MgO-based system, the LSMO peak is significantly shifted from the substrate due to the larger mismatch ($\sim$8.1 $\%$), suggesting partial strain relaxation\cite{Yadav2025, Lan2021,Lan2024}.\\
A similar behavior is observed for the Py layer, where the peak lies closer to LAO but remains well separated from MgO, reflecting the difference in lattice mismatch ($\sim$6$\%$ for LAO and $\sim$16$\%$ for MgO). The overall strain state, extracted from the RSM analysis (Table \ref{tab:table1}), indicates that in the Py/BTO/LSMO/LAO heterostructure both LSMO and BTO layers are under in-plane compressive strain, while the Py layer has in-plane tensile strain. In contrast, for Py/BTO/LSMO/MgO, LSMO undergoes in-plane tensile strain, whereas BTO remains compressed, with Py again exhibiting a in-plane tensile strain. Such differences in strain accommodation can lead to a more pronounced tetragonal distortion in BTO and LSMO for the LAO-based sample (Table \ref{tab:table1}). The compressive strain in BTO layer for both samples promotes out-of-plane elongation of unit cell, stabilizing a tetragonal distortion in BTO and favoring predominantly out-of-plane ferroelectric polarization\cite{Kim2022, AHN2015168}. In comparison, the Py layer shows relatively enhanced distortion in the MgO-based heterostructure. The in-plane and out-of-plane lattice parameters extracted from the RSM analysis (Fig. \ref{fig_xrd}(c,d)) are summarized in Table \ref{tab:table1} that matches well with those obtained from RHEED measurements. Further confirmation of epitaxy was done by XRD $\theta - 2\theta$, phi scan ($\phi$), and rocking curve measurement shown in the supplementary information section 1.
\begin{table*}[bt]
\centering
\setlength{\tabcolsep}{10pt}
\renewcommand{\arraystretch}{1.2}
\caption{\label{tab:table1}%
Out-of-plane (c) and in-plane (a) lattice parameters, corresponding perpendicular ($\epsilon_{\bot}$) and parallel epitaxial strains ($\epsilon_{\parallel}$), and tetragonal distortion $( \frac{c }{a} -1 )$ for Py/BTO/LSMO/LAO and Py/BTO/LSMO/MgO heterostructures.}
\begin{threeparttable}
\begin{tabular}{lccccc}
\toprule
\thead{Sample} 
& \thead{$c$ (\AA)} 
& \thead{$a$ (\AA)} 
& \thead{$\epsilon_{\bot}$ (\%)} 
& \thead{$\epsilon_{\parallel}$ (\%)} 
& \thead{Tetragonal distortion} \\
& & 
& $\left(\frac{c}{a_{0}}\right)-1$ 
& $\left(\frac{a}{a_{0}}\right)-1$ 
& $\left(\frac{c}{a}\right)-1$ \\
\midrule
\textbf{Py/BTO/LSMO/LAO}\\
Py layer & 3.52 ($\pm$0.07) & 3.60 ($\pm$0.06)  & -0.84 & 1.4 & - 0.023\\
BTO layer & 4.09 ($\pm$0.06) & 3.94 ($\pm$0.04)  & 2.51 & -1.25 &  0.038\\
LSMO layer & 3.99 ($\pm$0.04) & 3.81 ($\pm$0.09)  & 3.1 & -1.55 &  0.047\\
\textbf{Py/BTO/LSMO/MgO}\\
Py layer & 3.47 ($\pm$0.06) & 3.62 ($\pm$0.05)  & -2.25 & 1.97 & - 0.043\\
BTO layer & 4.06 ($\pm$0.03) & 3.97 ($\pm$0.07)  & 1.75 & -0.5 &  0.022\\
LSMO layer & 3.82 ($\pm$0.04) & 3.98 ($\pm$0.09)  & -1.29 & 2.84 & -0.04\\
\bottomrule
\end{tabular}
\begin{tablenotes}
\item Note: $a_0$ is the bulk lattice parameter of respective layers (Py, BTO and LSMO).
\end{tablenotes}
\end{threeparttable}
\end{table*}

\begin{figure*}[h]
 \centering
    \includegraphics[width=2\columnwidth]{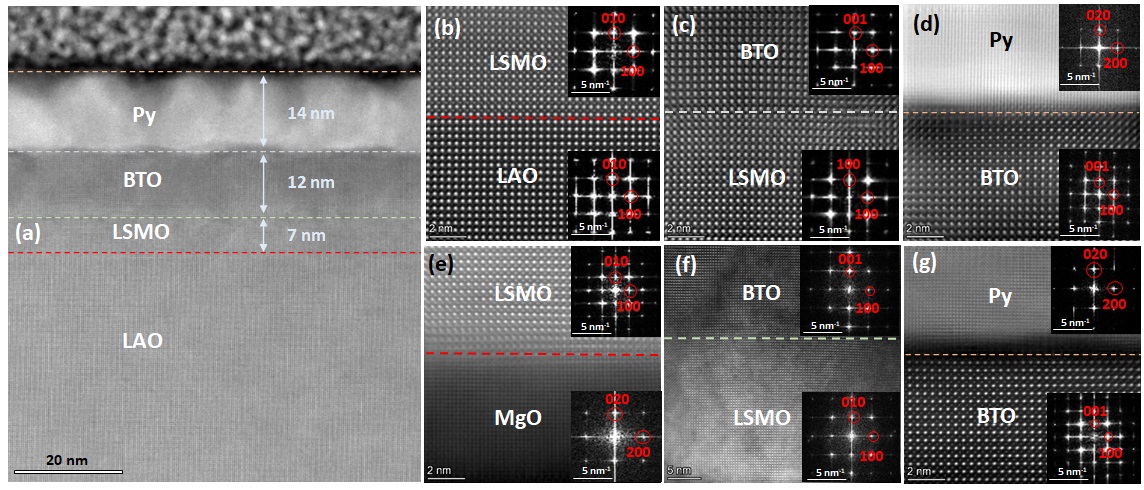}
    \caption{(a) HAADF-STEM cross-sectional image of the Py/BTO/LSMO/LAO heterostructure. Atomic-resolution HAADF images of the interfacial regions: (b) LSMO/LAO (Insets: FFT of LSMO, top-right; FFT of LAO, bottom-right); (c) BTO/LSMO (Insets: FFT of BTO, top-right; FFT of LSMO, bottom-right); and (d) Py/BTO (Insets: FFT of Py, top-right; FFT of BTO, bottom-right). Corresponding atomic-resolution HAADF images for the Py/BTO/LSMO/MgO heterostructure are shown in (e) LSMO/MgO (Insets: FFT of LSMO, top-right; FFT of MgO, bottom-right); (f) BTO/LSMO (Insets: FFT of BTO, top-right; FFT of LSMO, bottom-right); and (g) Py/BTO (Insets: FFT of Py, top-right; FFT of BTO, bottom-right).}
    \label{fig_TEM_images}
\end{figure*} 
Figure \ref{fig_TEM_images} shows the cross-sectional microstructure and epitaxial relationships within the multilayered Py/BTO/LSMO/LAO and Py/BTO/LSMO/MgO heterostructures, as revealed by atomic-resolution HAADF-STEM imaging. The cross-sectional HAADF-STEM image in Fig. \ref{fig_TEM_images} (a) of the entire Py/BTO/LSMO/LAO heterostructure highlights well-defined layering with sharp and flat interfaces and controlled thicknesses (LSMO $\sim$ 7 nm, BTO $\sim$ 12 nm, Py $\sim$ 14 nm), indicating high-quality epitaxial growth across the interfaces. The atomic-resolution HAADF-STEM image of the LSMO/LAO interface (Fig. \ref{fig_TEM_images} b) demonstrates an atomically sharp interface with excellent lattice continuity. FFTs confirm epitaxial alignment (010)LSMO $\parallel$ (010)LAO and (100)LSMO $\parallel$ (100)LAO, indicative of coherent strain transfer from the substrate. Compared to LAO, the LSMO/MgO interface on the MgO substrate (Fig. \ref{fig_TEM_images} e) induces distinct strain states due to larger lattice mismatch. FFTs confirm epitaxial alignment (010)LSMO $\parallel$ (020)MgO, highlighting substrate-dependent lattice matching. This difference in strain accommodation between LAO and MgO substrates is pivotal for tailoring anisotropy and transport properties. The BTO/LSMO interfaces (Fig. \ref{fig_TEM_images} c, f) show sharp lattice fringes but exhibit localized dislocations, suggesting partial strain relaxation. FFTs reveal epitaxial matching (001)BTO $\parallel$ (010)LSMO and (100)BTO $\parallel$ (100)LSMO in both the heterostructures. These dislocations are critical, as they may mediate polarization-driven charge redistribution and influence magnetoelectric coupling. Finally, in the Py/BTO interfaces (Figs. Fig. \ref{fig_TEM_images} d, g) dislocations are more pronounced, reflecting semi-coherent interfaces due to large lattice mismatch between metallic Py and oxide BTO. FFTs indicate orientation relationships (020)Py $\parallel$ (001)BTO and (200)Py $\parallel$ (100)BTO. Such semi-coherency is expected to impact spin transport and Rashba-type spin–orbit interactions at the metal–oxide boundary. Overall, the figure underscores the hierarchy of interface quality: oxide–oxide interfaces (LSMO/LAO, BTO/LSMO) exhibit high coherence with controlled strain relaxation, while the metal–oxide interface (Py/BTO) is semi-coherent. These structural characteristics may directly govern emergent phenomena such as strain-mediated ferroelectric–magnetic coupling, polarization-controlled charge accumulation, and spin-dependent transport.

\subsection{Magnetization and Magneto-transport}
\begin{figure*}[h]
 \centering
    \includegraphics[width=2\columnwidth]{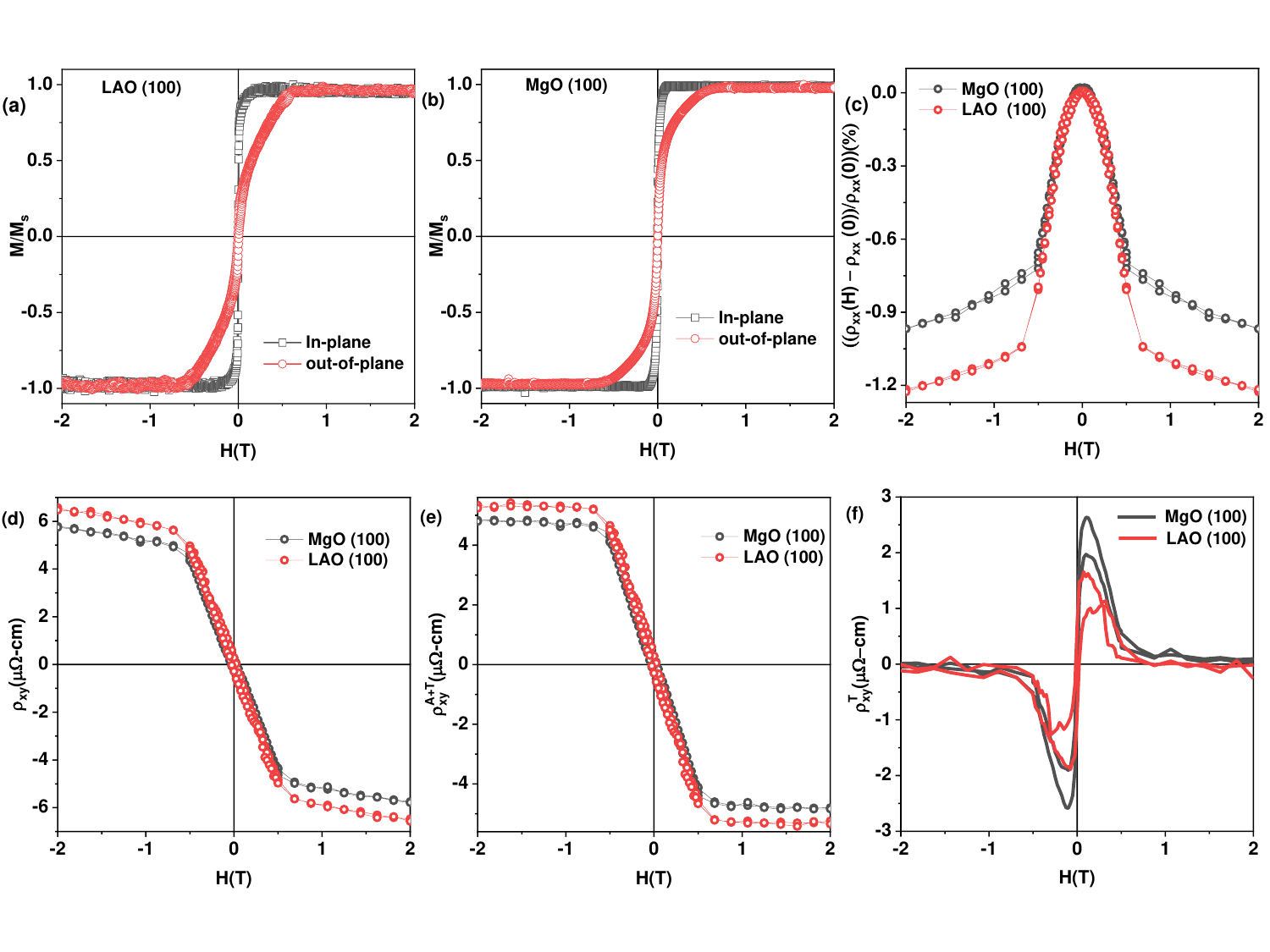}
    \caption{ Room-temperature magnetic and magnetotransport properties of Py/BTO/LSMO heterostructures grown on LAO (100) and MgO (100) substrates. The Py/BTO/LSMO/LAO and Py/BTO/LSMO/MgO samples are denoted as LAO (100) and MgO (100), respectively, in the figures. (a,b) Normalized magnetization curves for in-plane (black) and out-of-plane (red) field orientations. (c) Magnetoresistance of Py/BTO/LSMO/LAO (red) and Py/BTO/LSMO/MgO (black). (d) Magnetic field dependence of the total Hall resistivity. (e,f) Extracted anomalous and topological Hall resistivity components for the respective heterostructures. Data are displayed within ±2 T for clarity, measurements were performed up to 5 T }
    \label{fig_mag_and_Hall_MR}
\end{figure*} 
To investigate magnetic anisotropy in the heterostructures, the magnetization vs magnetic field M(H) curves were measured for the Py/BTO/LSMO/LAO and Py/BTO/LSMO/MgO heterostructures under the application of both in-plane and out-of-plane magnetic fields up to 5 T (shown upto $\pm$2 T for clarity,) as shown in Figs. \ref{fig_mag_and_Hall_MR}(a) and (b), where the Py/BTO/LSMO/LAO and Py/BTO/LSMO/MgO heterostructures are denoted as LAO (100) and MgO (100), respectively. From Fig.  \ref{fig_mag_and_Hall_MR}(a and b) it is evident that both LAO and MgO based heterostructures  exhibits distinct magnetic anisotropies. From Fig.  \ref{fig_mag_and_Hall_MR}(a) it observed that the LAO based heterostructure shows a uniaxial anisotropy, characterized by a relatively high in-plane squareness (M$_{r_\parallel}$/M$_{s_\parallel}$ $\sim$ 0.21) and low coercive field (H$_{C_\parallel}$ $\sim$ 11.2\ $\text{Oe}$, whereas the out-of-plane loop shows reduced squareness (M$_{r_\bot}$/M$_{s_\bot}$ $\sim$ 0.09) and significantly higher coercivity (H$_{C_\bot}$ $\sim$ 64.8\ \text{Oe}). In comparison, the MgO-based heterostructure also exhibits in-plane easy-axis behavior but with enhanced magnetic hardness, reflected by a higher in-plane squareness (M$_{r_\parallel}$/M$_{s_\parallel}$ $\sim$0.28) and larger coercivity (H$_{C_\parallel}$ $\sim$ 41.1 Oe). The corresponding out-of-plane response shows very low squareness (M$_{r_\bot}$/M$_{s_\bot}$ $\sim$ 0.02) and comparatively smaller coercivity (H$_{C_\bot}$ $\sim$12.2 Oe). The extracted uniaxial anisotropy constant (K$_u$)\cite{Dhakal2010} (see supplementary section 2 for more details) further confirms that the Py/BTO/LSMO/MgO heterostructure exhibits stronger magnetic anisotropy than Py/BTO/LSMO/LAO shown in Table 2. \\
\begin{table*}[bt]
\centering
\setlength{\tabcolsep}{10pt}
\renewcommand{\arraystretch}{1.2}
\caption{\label{tab:table2}%
Room-temperature in-plane squareness (M$_{r_{\parallel}}$/M$_{s_{\parallel}}$) and coercive field (H$_{c_{\parallel}}$), along with out-of-plane squareness (M$_{r_{\bot}}$/M$_{s_{\bot}}$) and coercive field (H$_{c_\bot}$), as well as the uniaxial anisotropy constant (K$_u$), extracted from magnetic hysteresis loops for Py/BTO/LSMO/LAO and Py/BTO/LSMO/MgO heterostructures.}
\begin{threeparttable}
\begin{tabular}{lccccc}
\toprule
\thead{Sample} 
& \thead{$M_{r_{\parallel}}/M_{s_{\parallel}}$} 
& \thead{$H_{c_{\parallel}}$} 
& \thead{$M_{r_{\bot}}/M_{s_{\bot}}$} 
& \thead{$H_{c_\bot}$} 
& \thead{$K_u(erg/cm^3)$} \\
\midrule
Py/BTO/LSMO/LAO & 0.21 & 11.2  & 0.1 & 64.8 & 1$\times10^6$\\
Py/BTO/LSMO/MgO & 0.28 & 41.1  & 0.02 & 12.2 & 2.5$\times10^6$\\
\bottomrule
\end{tabular}
\end{threeparttable}
\end{table*}
This enhancement can be attributed to two key factors. First, the larger strain in the Py layer when grown on BTO/LSMO/MgO in comparison to BTO/LSMO/MgO (Table \ref{tab:table1}), which modifies the magnetoelastic energy, thereby enhancing the magnetic anisotropy \cite{Yadav2025, Mondal2025}. Second, the enhanced magnetic anisotropy can be further understood by considering the strain-driven change of orbital occupancy and its influence on ferroelectric polarization at the BTO/LSMO interface. In the MgO-based heterostructure, the large lattice mismatch imposes an in-plane tensile strain on LSMO, which favors the occupation of the in-plane d$_{x^2-y^2}$ orbitals of Mn ions due to Jahn–Teller distortion. This orbital configuration promotes hole accumulation at the interface, leading to an upward polarization in the adjacent BTO layer. In contrast, for the LAO-based heterostructure, the in-plane compressive strain stabilizes the out-of-plane d$_{z^2-r^2}$ orbital occupancy, resulting in hole depletion at the interface and consequently favoring a downward polarization in BTO \cite{Chen2014, Hwang2012, Chen2012}.
Moreover, the BTO layer in the LAO-based system exhibits a larger tetragonal distortion compared to the MgO-based heterostructure, shows stronger compressive strain and stronger out-of-plane polarization\cite{Li2022}. The direction of polarization plays a crucial role in determining the interfacial magnetic properties. As demonstrated in ferroelectric/ferromagnetic heterostructures, polarization reversal modifies the atomic displacements at the interface, thereby altering orbital hybridization and spin-dependent charge transfer between the transition metal ions \cite{Begue2025_adv_Mater,Chatterjee2024,Li2023,Chen2020,Lu2012, Chakhalian2007, Duan2006}. In the present case, the polarization direction governs the strength and symmetry of hybridization between Py 3d states and Ti/Mn orbitals, which in turn modulates the magnetoelastic and magnetocrystalline anisotropy. Hence, the combined effects of strain-driven orbital ordering and polarization-controlled interfacial coupling account for the observed variation in magnetic anisotropy between the MgO- and LAO-based heterostructures.\\
The magnetotransport response demonstrates a pronounced enhancement in magnetoresistance for LAO based heterostructure relative to the MgO based heterostructure, as shown in Fig. \ref{fig_mag_and_Hall_MR} (c). This difference can be correlated with the distinct magnetic anisotropy of the two heterostructures. The higher out-of-plane coercivity and squareness (M$_r$/M$_s$) in the LAO based heterostructure indicate stronger domain wall pinning and a more stable magnetic configuration along the hard axis compared to the MgO-based sample. In this case, under an applied out-of-plane magnetic field, this stronger anisotropic constraint results in a more gradual rotation of magnetization from the in-plane easy axis toward the out-of-plane direction, producing a larger variation in spin-dependent scattering and thereby increasing the magnetoresistive response \cite{Conte2018,Meng2016}. By contrast, the MgO based heterostructure, with its lower out-of-plane coercivity, allows faster alignment of magnetization along the applied field direction, leading to a smaller change in resistivity during field sweeping. The structural distortion of the BTO layer also contributes to this behavior. The BTO in the LAO-based heterostructure exhibits a larger tetragonal distortion (c/a$\sim$1.04), indicating stronger spontaneous polarization than in the MgO-based system \cite{Li2022}. This enhanced polarization can generate a larger interfacial electric field, which strengthens Rashba-type spin–orbit coupling at the interface. Such interfacial spin–orbit coupling can increase spin-dependent scattering and modify charge transport, thereby contributing to the larger percentage change in magnetoresistance \cite{Narayanapillai2017, yamada2021, Shen2023}. Therefore, the combined effects of stronger magnetic pinning and polarization-driven interfacial coupling can give rise to the enhanced magnetoresistance observed in the LAO based heterostructure.\\

A similar trend is observed in the Hall transport measurements, where the magnetic-field dependence of the total Hall resistivity ($\rho_{xy}$) at room temperature shows a larger magnitude for the LAO based heterostructure compared to the MgO-based system, as shown in Fig \ref{fig_mag_and_Hall_MR} (d). In ferromagnetic thin films, the total Hall resistivity can be expressed as the sum of ordinary, anomalous, and topological contributions, given by \cite{Yanda2025, Yadav2025, Mondal2025, Lone2024, Ma2022, He2018, Soumyanarayanan2017}
\begin{equation}
   \rho_{xy}(H)= \rho_{xy}^{0}+\rho_{xy}^{A}+\rho_{xy}^{T}= R_0 H+ R_SM+\rho_{xy}^{T}
    \label{eqTHE}
\end{equation}
where the $\rho_{xy}^{0}$, $\rho_{xy}^{A}$ and $\rho_{xy}^{T}$ corresponds to ordinary, anomalous, and topological Hall resistivity contributions, respectively. Here, $R_{0}$ is the ordinary Hall coefficient, R$_S$ is the anomalous Hall coefficient and M represents the out-of-plane magnetization of the films. The carrier concentration was obtained from the ordinary Hall coefficient ($R_{0}$), by linear fitting of the high-field region, where the Hall response is dominated by the ordinary contribution. The negative slope of the high-field linear region confirms electron-type charge carriers in both heterostructures. The extracted carrier densities are $1.03(\pm 0.01) \times 10^{21}\ \text{cm}^{-3}$ for the LAO based sample and $1.3(\pm 0.01) \times 10^{21}\ \text{cm}^{-3}$ for the MgO based sample at room temperature, indicates comparable carrier concentrations in both heterostructures. To further analyze the Hall response, the ordinary Hall contribution was subtracted, and the resulting combined anomalous and topological Hall resistivity ($\rho_{xy}^{A+T}$ = $\rho_{xy}$ - $\rho_{xy}^{0}$) is presented in Fig. \ref{fig_mag_and_Hall_MR} (e). A clear enhancement in the Hall response is observed for the LAO based heterostructure compared to the MgO-based system. This increased magnitude is consistent with the magnetoresistance behavior and can be attributed to the differences in magnetic anisotropy and interfacial spin–orbit coupling between the two heterostructures.
In the LAO-based system, the larger M$_{r_\bot}$/M$_{s_\bot}$ and higher out-of-plane coercivity promote a more gradual and stable magnetization evolution under applied field, which enhances the anomalous Hall contribution. Additionally, the larger tetragonal distortion of the BTO layer (c/a $\sim$ 1.04) indicates stronger ferroelectric polarization, leading to an enhanced interfacial electric field. This can induce stronger Rashba-type spin–orbit coupling, which is known to amplify the anomalous Hall effect through enhanced Berry curvature and spin-dependent scattering mechanisms \cite{Wei2025,Zhang2023}.
In contrast, the relatively smaller M$_{r_\bot}$/M$_{s_\bot}$, out-of-plane coercivity and reduced polarization in the MgO-based heterostructure result in a smaller Hall response. Therefore, the combined effects of anisotropy-driven magnetization dynamics and polarization-enhanced interfacial spin–orbit coupling can accounts for the larger anomalous Hall contribution observed in the LAO based heterostructure.\\
However the topological Hall resistivity ($\rho^T_{xy}$), obtained by subtracting the anomalous contribution from the combined signal,  i.e.,$ \rho_{xy}^{T} = \rho_{xy}^{A+T} - \rho_{xy}^{A}$ exhibits a larger magnitude in the MgO based heterostructure compared to the LAO based system, as shown in Fig. \ref{fig_mag_and_Hall_MR} (f). This enhancement of the topological Hall signal in the MgO-based heterostructure can be understood from the strain-driven modification of interfacial polarization and orbital reconstruction. In the MgO based system, the tensile strain imposed on LSMO stabilizes the in-plane d$_{x^2-y^2}$ orbitals via Jahn–Teller distortion, promoting hole accumulation at the interface. This favors an upward polarization of the BTO layer. In contrast, the compressive strain in the LAO based heterostructure stabilizes the out-of-plane d$_{3z^2-r^2}$ orbitals, leading to hole depletion and a polarization in the downward direction \cite{Chen2014,Hwang2012, Chen2012}. Such strain-dependent polarization states induce distinct polar distortions within the interfacial MnO$_2$ layer in LSMO, where the displacement amplitude is generally larger under tensile strain and reduced under compressive strain \cite{Chen2014, Chatterjee2024}. Furthermore, in the case of upward polarization (MgO-based system), the Ti ions shift toward the top interface, enhancing the overlap between transition-metal 3d orbitals across the interface. This leads to stronger hybridization between Py 3d states and Ti/Mn orbitals, modifying the interfacial electronic structure and spin polarization. As a consequence, the redistribution of spin-dependent electronic states can enhance interfacial magnetic moments and promote stronger ferroelectric–magnetic proximity effects \cite{Chen2014, Chatterjee2024, Duan2006, Grisolia2016}. Such interfacial coupling is known to play a key role in stabilizing non-collinear spin textures through enhanced spin–orbit interactions and interfacial Dzyaloshinskii–Moriya interaction (DMI)\cite{Yao2022,Ohuchi2018, Yoo2021}.
In contrast, the downward polarization in the LAO-based heterostructure results in weaker interfacial hybridization and reduced asymmetry, which suppresses the formation of chiral spin configurations. Therefore, despite exhibiting a larger anomalous Hall effect, the LAO-based system shows a comparatively weaker topological Hall response. These results highlight that the combined effects of strain-driven orbital ordering, polarization direction, and interfacial spin–orbit coupling govern the stabilization of non-trivial spin textures, leading to an enhanced topological Hall effect in the MgO based heterostructure.

\subsection{ Ferroelectric Property of BTO}
\begin{figure*}[h]
 \centering
    \includegraphics[width=2\columnwidth]{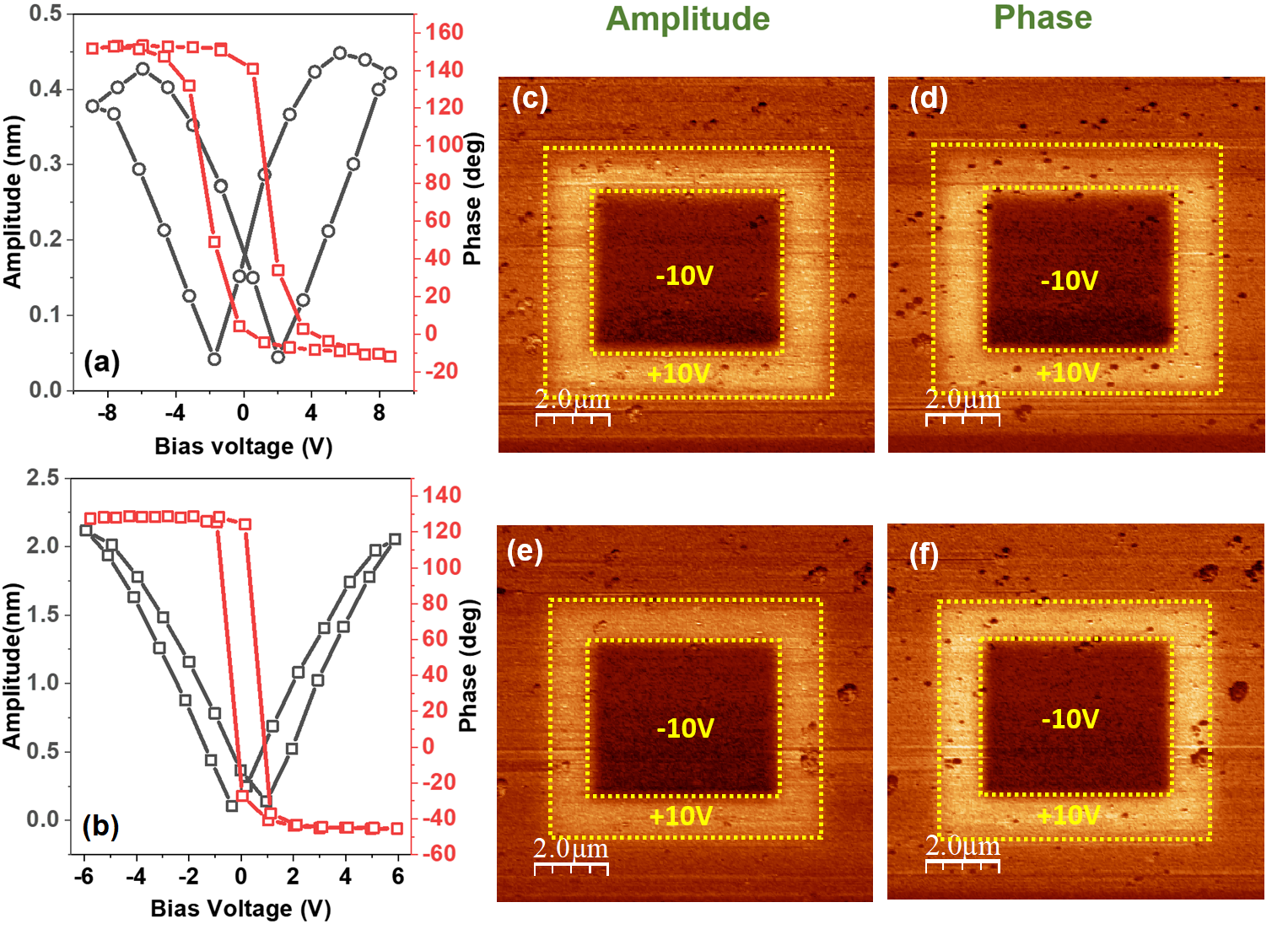}
    \caption{PFM characterization of the ferroelectric BTO layer in BTO/LSMO heterostructures. (a,b) Amplitude (black) butterfly loops and phase (red) hysteresis loops for BTO films on MgO (100) and LAO (100) substrates, respectively. (c,d) Out-of-plane amplitude and phase images of BTO/LSMO/MgO after $-$10 V (center) and + 10 V (outer) poling, respectively. (e,f) Corresponding out-of-plane amplitude and phase images for BTO/LSMO/LAO. The scale bar in all images represents 2 $\mu$m. }
    \label{fig_BTO_PFM}
\end{figure*} 
The  ferroelectric order of the BTO layer was probed using piezoresponse force microscopy (PFM) as shown in Fig. \ref{fig_BTO_PFM}.  In Figs. \ref{fig_BTO_PFM}(a–b) we apply DC biases ($\pm$8 V for MgO, $\pm$ 6 V for LAO) and measure the local PFM phase and amplitude.  Both samples exhibit clear 180$^0$ phase hysteresis loops (red curves) and characteristic butterfly-shaped amplitude loops (black curves), demonstrating electric-field–driven polarization reversal in BTO.  Notably, the coercive bias for BTO is larger on MgO ($\sim$ 1.5 V) than on LAO ($\sim$ 0.5 V) substrates,  which can be attributed due to strain and interfaces influencing the ferroelectric property \cite{JinHu2016, Kelley2020, Li2024, Chen2013}. To further verify polarization reversal and retention, local poling experiments were performed in which a 2$\times$2 $\mu$m$^2$ central region was written with a –10 V bias (upward polarization), while the surrounding 5$\times$5 $\mu$m$^2$ area was poled using +10 V (downward polarization). Subsequent PFM imaging over a 10$\times$10 $\mu$m$^2$ area using a small AC voltage (5 V) reveals distinct “box-in-box” contrast in both phase and amplitude images (Fig. \ref{fig_BTO_PFM} c–f), where the central region exhibits a clear 180$^0$ -phase contrast relative to the surrounding area, indicating complete and stable 180$^0$ polarization reversal.  Overall, these results demonstrate that the BTO layers on both MgO- and LAO-based heterostructures exhibit stable, switchable, and well-retained ferroelectric polarization \cite{JinHu2016, Kelley2020, Li2024, Chen2013}. 
\begin{figure*}[h]
 \centering
    \includegraphics[width=2\columnwidth]{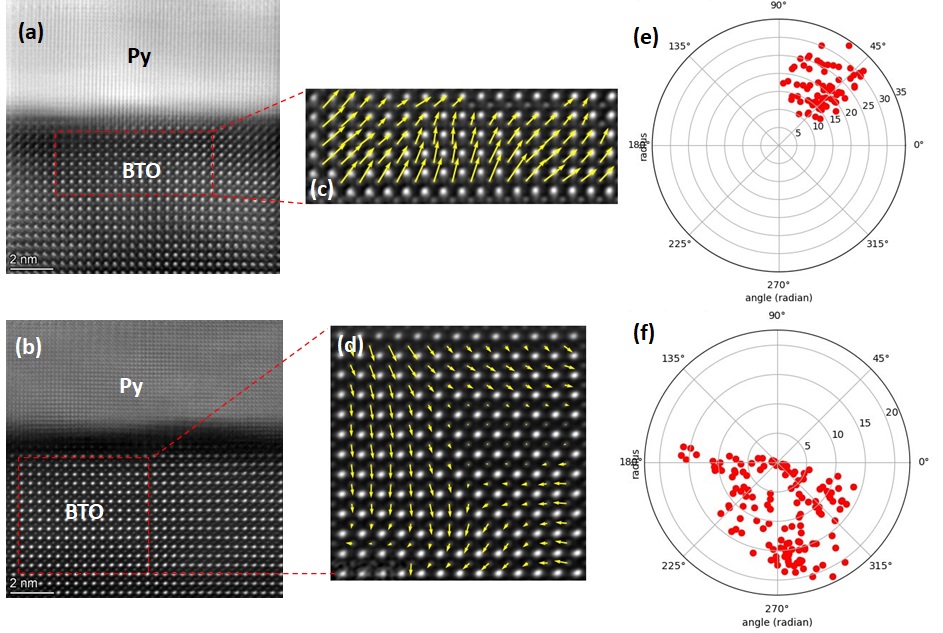}
    \caption{ Cross-sectional HAADF-STEM imaging and polarization mapping at Py/BTO interfaces. (a, b) High-resolution HAADF-STEM images of the Py/BTO interfaces, with red rectangular boxes marking regions selected for polarization analysis. (c, d) Enlarged views of the boxed regions, where yellow arrows indicate local polarization vectors derived from atomic displacements. (e, f) Polar plots of polarization directions in BTO adjacent to the Py interface: in the Py/BTO/LSMO/LAO heterostructure, polarization vectors are predominantly oriented between 45° and 135°, consistent with upward polarization alignment; in contrast, the Py/BTO/LSMO/MgO heterostructure exhibits polarization vectors mainly between 180° and 270°, corresponding to downward polarization orientation.}
    \label{fig_BTO_TEM_polarisationmapping}
\end{figure*} 
Figure \ref{fig_BTO_TEM_polarisationmapping} presents cross-sectional HAADF-STEM images and polarization mapping at the Py/BTO interfaces, offering direct insight into how interfacial structure governs ferroelectric behavior in complex heterostructures. Figures \ref{fig_BTO_TEM_polarisationmapping} (a,b) show high-resolution HAADF-STEM images of the Py/BTO interfaces, with red rectangular boxes marking the regions analysed for polarization for the Py/BTO/LSMO/LAO and Py/BTO/LSMO/MgO heterostructures, respectively. The magnified views in Figs. \ref{fig_BTO_TEM_polarisationmapping} (c,d) highlight atomic displacements within the BTO lattice, with yellow arrows indicating local polarization vectors. These observations confirm that ferroelectric distortions persist even in close proximity to the metallic Py layer, underscoring the robustness of BTO polarization at these semi-coherent metal–oxide boundaries. The corresponding polar plots in Figs. \ref{fig_BTO_TEM_polarisationmapping} (e,f) quantify the distribution of polarization directions near the Py interface. For the Py/BTO/LSMO/LAO heterostructure, polarization vectors predominantly align between 45° and 135°, suggesting an upward polarization orientation. In contrast, the Py/BTO/LSMO/MgO heterostructure exhibits polarization vectors concentrated between 180° and 270°, indicating a downward polarization orientation. This substrate-dependent polarization alignment highlights the critical role of strain states and interfacial dislocations in dictating ferroelectric domain stability. The angular spread further reflects localized strain accommodation, which may enable tunable charge accumulation and Rashba-type spin–orbit coupling at the Py/BTO interface. Collectively, these results demonstrate that while the oxide–oxide interfaces maintain coherent epitaxy, the metal–oxide Py/BTO interface remains semi-coherent yet capable of sustaining switchable ferroelectric polarization. The persistence and orientation of polarization at the nanoscale provide a structural foundation for engineering strain-mediated magnetoelectric coupling, polarization-controlled spin transport, and substrate-dependent Hall responses in multifunctional heterostructures.
\subsection{Electric field control of AHE}
\begin{figure*}[h]
 \centering
    \includegraphics[width=2\columnwidth]{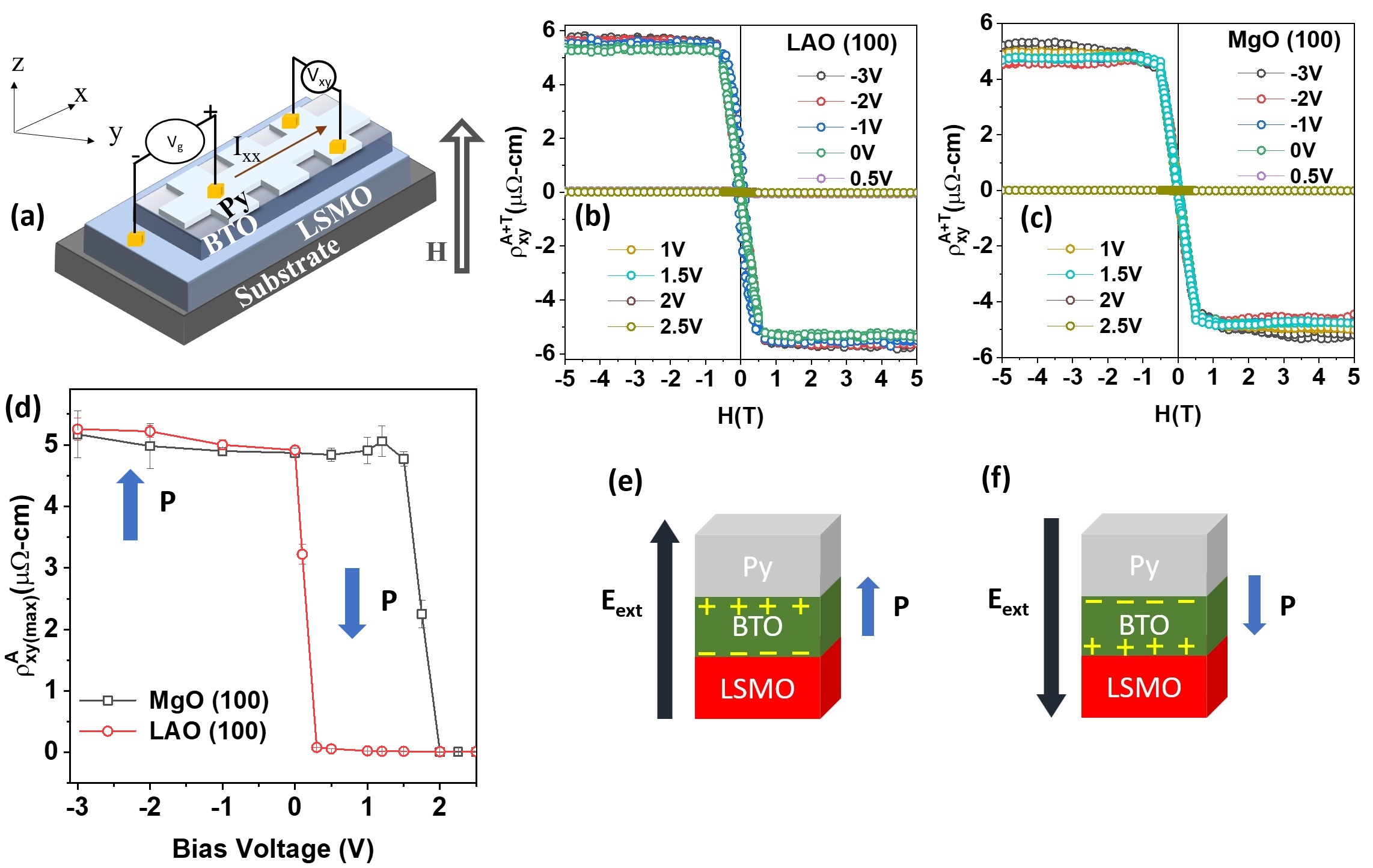}
    \caption{Room-temperature electric-field-modulated Hall resistivity in Py/BTO/LSMO heterostructures. The Py/BTO/LSMO/LAO and Py/BTO/LSMO/MgO samples are denoted as LAO (100) and MgO (100), respectively, in the figures. (a) Schematic of the device geometry and measurement configuration, shows the applied gate voltage (V$_g$) across the BTO layer and generated transverse (V$_{xy}$) Hall voltage measurements. (b,c) Magnetic field dependence Anomalous $+$ topological Hall ($\rho^{A+T}_{xy}$ ) contributions for LAO and MgO heterostructures, respectively. (d) Maximum anomalous Hall resistivity $(\rho_{xy_{max}}^{A}$) as a function of applied bias voltage for the LAO (red) and MgO (black) based heterostructures. Schematic  in (e,f) show polarization-controlled interfacial charge redistribution under positive and negative external electric fields (E$_{ext}$)}
    \label{voltagedependent}
\end{figure*} 
Building on the ferroelectric switching behavior discussed above, the influence of electric-field control on Hall properties is examined in Fig. \ref{voltagedependent}. The measurement configuration shown in Fig. \ref{voltagedependent} (a), a DC bias voltage is applied across the BTO layer by contacting the top Py electrode and the bottom LSMO layer, enabling direct control of the ferroelectric polarization through the applied electric field. Simultaneously, a longitudinal current is passed along the in-plane direction the generated transverse Hall voltage is recorded under an applied out-of-plane magnetic field. The magnetic-field dependence of anomalous plus topological contribution ($\rho^{A+T}_{xy}$) under different applied bias voltages from - 3 V to 2.5 V for the LAO and MgO based heterostructure are shown in Fig. \ref{voltagedependent}(b–c). A pronounced change of the Hall response with applied bias is clearly observed in both systems. Notably, under positive bias (corresponding to an electric field directed from Py toward LSMO), a significant change in the Hall resistivity occurs at relatively low voltages ($\sim$0.5 V for the LAO-based heterostructure and $\sim$2 V for the MgO-based system). In contrast, reversing the polarity up to higher voltages results in only a weak variation in the Hall signal.\\
The maximum variation in anomalous Hall resistivity $(\rho_{xy_{max}}^{A}$) as a function of applied bias voltage is summarized in Fig. \ref{voltagedependent} (d), while the corresponding polarization configurations and carrier change mechanism are schematically shown in Fig. \ref{voltagedependent} (e–f). The pronounced voltage-dependent change of $\rho_{xy}^{A}$ can be understood in terms of ferroelectric polarization-controlled carrier redistribution, interfacial orbital reconstruction, and electric-field-driven spin–orbit coupling at the Py/BTO/LSMO interface. Under a positive electric field (E$_{\mathrm{ext}}$), applied from the Py top layer toward the LSMO bottom electrode, the ferroelectric polarization of the BTO layer switches downwards, whereas reversal of the electric-field polarity drives the polarization upwards. Below the coercive voltage of BTO, the polarization state remains nearly unchanged (shown in Fig. \ref{fig_BTO_PFM}) and therefore only a weak variation in $\rho_{xy}^{A}$ is observed. However, as the applied voltage approaches the coercive bias ($\sim$0.5 V for Py/BTO/LSMO/LAO and $\sim$2 V for Py/BTO/LSMO/MgO), abrupt polarization reversal occurs, producing strong bound charge accumulation at the interfaces. In the downward polarization state, negative bound charge accumulates near the Py/BTO interface while positive bound charge forms near the BTO/LSMO interface. This interfacial electrostatic reconstruction promotes electron accumulation in the interfacial metallic layers and substantially modifies the carrier density, as reflected in the extracted ordinary Hall coefficient analysis shown in supplementary section 3. Since the anomalous Hall resistivity is strongly dependent on carrier concentration and scattering asymmetry, the enhanced carrier density leads to a marked reduction in $\rho_{xy}^{A}$\cite{Kim2024, Liu2024}. In contrast, under negative bias the upward polarization produces positive bound charge near the Py/BTO interface and negative bound charge near the BTO/LSMO interface, resulting in comparatively weaker carrier change because the metallic Py layer efficiently screens the polarization charge. Consequently, the Hall response exhibits a pronounced asymmetry between positive and negative voltage sweeps.  This suggests the emergence of asymmetric band dispersion associated with Rashba-like spin splitting, \cite{Kim2024, Liu2024, Brehin2023}. Furthermore, the broken inversion symmetry associated with the polarized BTO layer introduces Rashba-type spin–orbit coupling at the interface \cite{Lin2018}. The interfacial electric field generated by ferroelectric polarization lifts spin degeneracy and produces asymmetric spin-split electronic bands, thereby modifying the anomalous Hall response through Berry-curvature-related mechanisms \cite{Lebedev2021, Kim2024, Liu2024, Ohuchi2018, Lin2018}. The Rashba interaction becomes stronger when the local electric field and orbital hybridization are enhanced, leading to larger spin-dependent scattering and enhanced anomalous Hall resistivity for one polarization state compared to the opposite orientation. Electrically controlled Hall change driven by polarization-dependent spin–orbit coupling and magnetic proximity effects has been reported in Co/Pt/PMN-PT\cite{Feng2020}, SRO/BTO heterostrucutre \cite{Ren2020, Liu2019}.
The strong asymmetry observed in the Hall response therefore can be originates from the combined effects of asymmetric polarization screening, carrier redistribution across the Py/BTO/LSMO interface, and polarization-dependent change of Rashba spin–orbit coupling and orbital hybridization \cite{Feng2020,Ren2020,Kim2024,Liu2024,Ohuchi2018,Chen2014,Lin2018}. These results further confirm that ferroelectric polarization in BTO not only controls interfacial charge accumulation but also directly tunes the spin-dependent electronic structure and magnetotransport properties of the heterostructure\\
Importantly, the voltage-dependent transport measurements demonstrate giant electric-field control of Hall transport, achieving nearly $\sim$93$\%$ change of anomalous Hall resistivity at exceptionally low operating voltages of only $\sim$0.5 V and $\sim$ 2 V for the LAO- and MgO-based heterostructures, respectively. Such a large room-temperature change in a metallic ferromagnetic heterostructure highlights the strong interplay among ferroelectric polarization, interfacial charge reconstruction, orbital hybridization, and Rashba-type spin–orbit coupling at the Py/BTO/LSMO interface. These results demonstrates the potential of these oxide heterostructures for realizing energy-efficient multifunctional spintronic devices and electrically tunable topological electronic systems.

\subsection{Results of the DFT calculations}
\begin{figure}[h]
		\centering
			\includegraphics[width=0.5\textwidth]{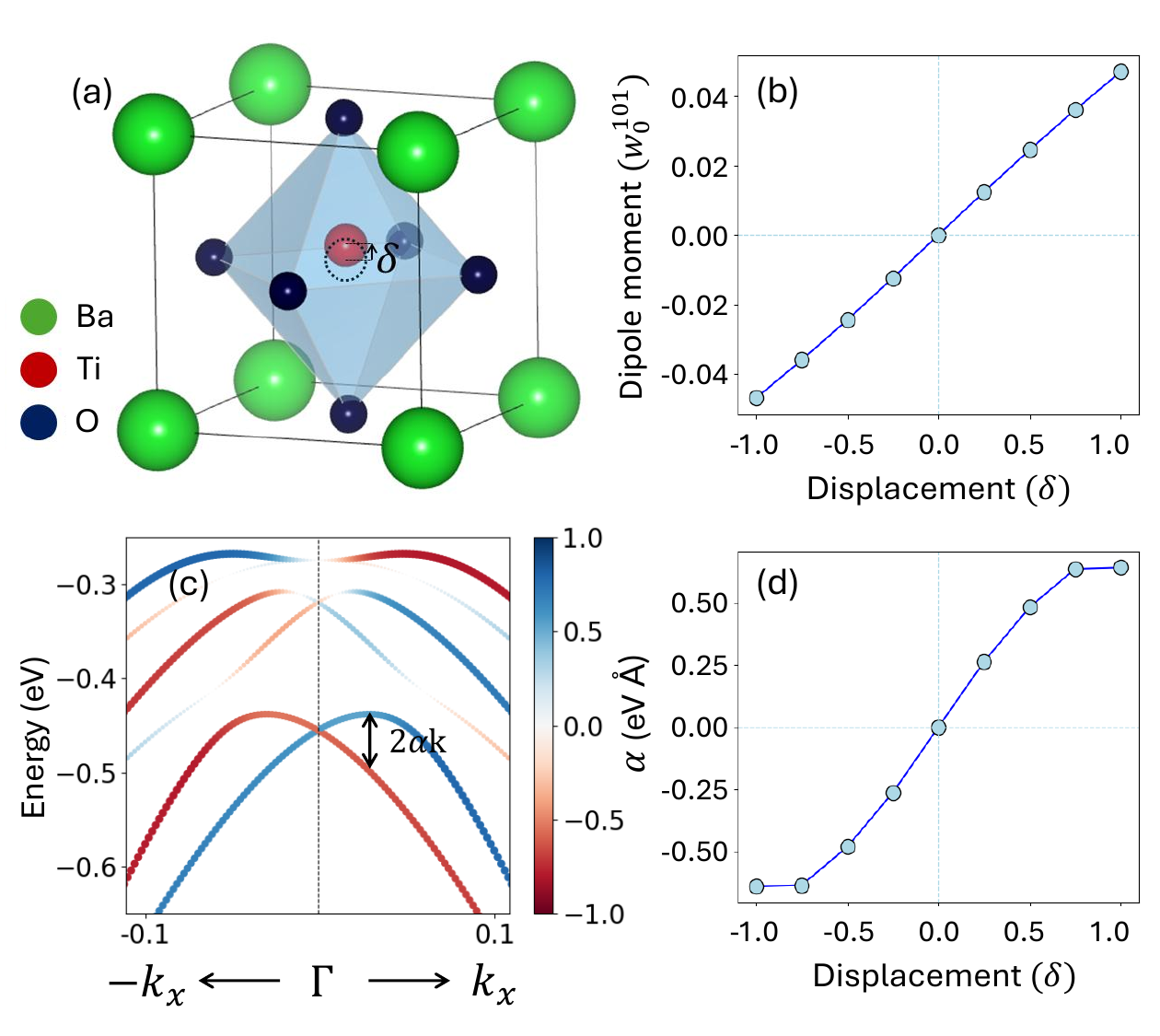} 
	\caption{(a) Crystal structure of the tetragonal BaTiO$_3$. Schematic illustration of the displacement $\delta$ of the Ti ion from the center of the oxygen cage. (b) The $w^{101}_{0}$ multipole component (in the units of electronic charge), corresponding to the $z$ component of the electric dipole moment, for different structures of BaTiO$_3$ with varying amplitudes of $\delta$ (see text for details). (c) Band structure of BaTiO$_3$ in the presence of spin-orbit coupling for the relaxed structure (see Table~\ref{tab1}). The Fermi energy is set at zero. The color bar shows the expectation value of the $y$-component of the spin, projected onto the band energies. (d) Values of the Rashba parameter $\alpha$ as a function of $\delta$.}
	\label{fig1}
\end{figure}

To gain further insight into the interplay between electric polarization and Rashba coupling, we also perform DFT calculations for BaTiO$_3$. The ferroelectric phase of BaTiO$_3$ has a tetragonal structure in which the atoms are off-centered along the $z$ axis. The schematic structure of the ferroelectric BaTiO$_3$, as shown in Fig \ref{fig1}a, depicts the shifting of the Ti$^{4+}$ ion in the upward direction with respect to the surrounding oxygen cage. The structure has the $P4mm$ symmetry that also allows the O and Ba atoms to undergo displacements along the $z$ direction, as evident from the relaxed atomic positions displayed in Table~\ref{tab1}. This off-centering of the atoms is responsible for the spontaneous polarization in the ferroelectric phase of BaTiO$_3$. From the listed relaxed atomic positions in Table~\ref{tab1}, we extract the atomic displacement $\delta$ for each ion from their centrosymmetric positions. Following this, we artificially construct several other tetragonal structures with the same lattice parameters, but varying the atomic displacements from -$\delta$ to $\delta$ at an interval of $\delta/4$. The structure corresponding to a negative $\delta$ represents the opposite ferroelectric domain with the reversed direction of atomic displacements.

\begin{table}[h]
\centering
\caption{Relaxed atomic positions of BaTiO$_3$.}
\begin{threeparttable}
\begin{tabular}{lcccc}
\toprule
\thead{Site} 
& \thead{Wyckoff position} & \thead{x} & \thead{y} & \thead{z}\\
\midrule
Ba & 1a & 0.0 & 0.0 & 0.001\\
Ti & 1b & 0.5 & 0.5 & 0.517\\
O  & 1b & 0.5 & 0.5 & 0.973\\
O  & 2c & 0.5 & 0.0 & 0.484\\
\bottomrule
\end{tabular}
\end{threeparttable}
\label{tab1}
\end{table}

To understand the effect of these atomic displacements on the spontaneous polarization of the ferroelectric BaTiO$_3$, we calculate the electric dipole moments at every atomic site for these different structures. Our calculations show that only the $z$ component of the dipole moment has nonzero values. Fig.~\ref{fig1}b shows the variation of the $z$ component of the dipole moments as a function of $\delta$. As seen from the plot, the magnitude of the electric polarization increases with increasing $\delta$. Additionally, as expected, the direction of polarization is reversed for the reversed atomic displacements, indicated by the opposite sign of $\delta$. \\
The broken inversion symmetry due to electric polarization along $\hat z$ in the presence of spin-orbit coupling gives rise to the Rashba-like spin-splitting in the band structure of BaTiO$_3$. The spin-orbit coupling is a relativistic effect in which an electron, moving around a nucleus, sees the electric field as a magnetic field in its rest frame. This magnetic field, in turn, interacts with the spin of the electron. In solids lacking an inversion symmetry \cite{rashba}, the Rashba effect is due to the electric field arising from a non-vanishing gradient of the crystalline electric potential as $V(r) \neq V(-r)$. This leads to the following spin-orbit coupling term in the low-energy Hamiltonian~\cite{rashba2},
\begin{eqnarray}
    H_R = \alpha \, ({\vec{\sigma}} \times {\vec{k}}) \cdot \hat{z} = \alpha \, (\sigma_x k_y - \sigma_y k_x)
    \label{eq1}.
\end{eqnarray}
Here, $\vec{k}$ is the crystal momentum, $\alpha$ and $\sigma$ are respectively the Rashba coupling coefficient and the Pauli spin operator. The Rashba term in Eq. (\ref{eq1}) lifts the Kramers' degeneracy and introduces a $k$ dependent spin splitting of electronic bands, as evident from the energy eigenvalues, $ E_{\pm} = \pm \alpha k$, where $k=\sqrt{k_x^2 + k_y^2}$.\\
The band structure of BaTiO$_3$ corresponding to the crystal structure given in Table \ref{tab1} in the presence of spin-orbit coupling around the $\Gamma$ point is shown in Fig.~\ref{fig1}c. A significant spin-splitting for the valence bands, resembling the Rashba spin-splitting, is evident from Fig.~\ref{fig1}c. By fitting the energy splitting, $\Delta E=E_\uparrow -E_\downarrow= 2\alpha k$, near the $\Gamma$ point, we extract the Rashba parameter $\alpha$. Here, $E_\uparrow$ and $E_\downarrow$ are the energies of spin-up and spin-down states, respectively. Our computed value is $\alpha$ = 0.64 eV{\AA}.\\
We note that the eigenstates for the Rashba Hamiltonian in Eq. (\ref{eq1}) are spin-polarized, with the following spin expectation values,
 \begin{eqnarray}
    \langle \sigma_x \rangle_{+} = \frac{k_y}{k}, 
    \quad 
    \langle \sigma_y \rangle_{+} = -\frac{k_x}{k},
    \quad
    \langle \sigma_z \rangle_{+} = 0,
    \end{eqnarray}
and
 \begin{eqnarray}
    \langle \sigma_x \rangle_{-} = -\frac{k_y}{k}, 
    \quad 
    \langle \sigma_y \rangle_{-} = \frac{k_x}{k},
    \quad
    \langle \sigma_z \rangle_{-} = 0.
    \end{eqnarray}
Thus, the spins are locked in the $(k_x,k_y)$ plane, leading to a head-to-tail spin texture. Consistent with this, we find that the bands along $k_x$ are spin polarized along $\hat y$ (see Fig.~\ref{fig1}c).\\
We further compute the band structures for the different crystal structures with varying $\delta$, discussed earlier, and extract the Rashba parameter $\alpha$ for each of these cases. The result of our calculation is shown in Fig. \ref{fig1}d. 
As seen from the plot, the magnitude of $\alpha$ reduces as the atomic displacement decreases. Since the atomic displacements govern the electric polarization, as depicted in Fig. \ref{fig1}b, this indicates the tuning of the Rashba spin-splitting by controlling the ferroelectric polarization of BaTiO$_3$. Interestingly, we note that the sign of $\alpha$ is also reversed for the structures having atomic displacements in the reversed direction, indicating the reversal of spin polarization of the split bands with the reversal of ferroelectric polarization. For $\delta=0$, the ferroelectric polarization is absent due to the presence of inversion symmetry, and, consequently, the Rashba spin-splitting disappears as well. This demonstrates that the ferroelectric polarization not only controls the magnitude but also the sign of the Rashba parameter $\alpha$ in ferroelectric BaTiO$_3$.
\section{Conclusion}
In summary, strain-engineered Py/BTO/LSMO heterostructures grown on MgO and LAO substrates exhibit strongly coupled structural, ferroelectric, magnetic, and spin-dependent transport properties governed by substrate-induced lattice distortion and interfacial polarization effects. Reciprocal space mapping and atomic-resolution HAADF-STEM imaging reveal distinct strain accommodation mechanisms in the two heterostructures, leading to different tetragonal distortions, orbital configurations, and polarization states in the BTO and LSMO. The MgO-based heterostructure exhibits enhanced in-plane magnetic anisotropy together with a significantly larger topological Hall effect, whereas the LAO-based system shows stronger anomalous Hall response and enhanced magnetoresistance. The enhanced topological Hall signal in the MgO-based heterostructure is attributed to tensile-strain-induced orbital reconstruction, polarization-directed carrier accumulation, and stronger ferroelectric–magnetic proximity coupling at the interface, which collectively stabilize nontrivial spin textures. In contrast, the larger anomalous Hall resistivity and magnetoresistance observed in the LAO-based heterostructure originate from enhanced out-of-plane magnetic anisotropy, stronger Rashba-type spin–orbit coupling, and polarization-driven interfacial spin-dependent scattering. Piezoresponse force microscopy and HAADF-STEM polarization mapping confirm robust and reversible ferroelectric switching in both heterostructures, while voltage-dependent transport measurements demonstrate giant electric-field control of Hall transport, achieving nearly $\sim$93$\%$ change of anomalous Hall resistivity at remarkably low operating voltages of only $\sim$0.5 V and $\sim$2 V through polarization-mediated carrier redistribution and interfacial spin–orbit interactions. The strong asymmetry in the electrically controlled Hall response further highlights the important role of polarization-dependent charge screening, orbital hybridization, and inversion-symmetry breaking at the interface. This is further corroborated by our theoretical findings, which demonstrate the tunability of Rashba spin splitting in BTO through control of the ferroelectric polarization. These findings establish a direct correlation among strain-controlled ferroelectric polarization, orbital hybridization, Berry-curvature-driven transport, and interfacial spin–orbit coupling in oxide heterostructures, providing an effective route towards low-power electrically tunable multifunctional spintronic devices.
\section{Methodology}
In this section, we describe the experimental methods and the computational methods employed in this work.
\subsection{Experimental methods}
High-purity (99.99\%) commercially bought targets of permalloy (Py, Ni$_{80}$Fe$_{20}$), La$_{0.7}$Sr$_{0.3}$MnO$_3$ (LSMO), and BaTiO$_3$ (BTO) (Kurt J. Lesker Company) were used in the thin film depositions.  Py/BTO/LSMO heterostructures, were fabricated on single-crystal MgO (100) and LaAlO$_3$ (LAO) (100) substrates by combining pulsed laser deposition (PLD; Excel Instruments system equipped with an in-situ reflection high energy electron diffraction (RHEED)) and DC magnetron sputtering (Minilab S80A, Moorfield Nanotechnology). Initially, epitaxial BTO/LSMO bilayers were deposited on 5 mm × 5 mm MgO and LAO substrates using the PLD technique. The LSMO and BTO targets were sequentially ablated using a KrF excimer laser ($\lambda$ = 248 nm) operated at a repetition rate of 5 Hz with a laser fluence of $\sim$ 3 J cm$^{-2}$. The deposition chamber was equipped with a multi-target carousel, enabling in situ growth of multilayers. During deposition, the target-to-substrate distance was maintained at $\sim$ 5 cm. Prior to film growth, the MgO and LAO substrates were thermally treated at 800 $^0C$ for 2 h under an oxygen partial pressure of 500 mTorr to ensure surface cleanliness and promote epitaxial growth. Subsequently, LSMO layers with a thickness of $\sim$ 7 nm were deposited at 750 $^0C$ under an oxygen pressure of 20 mTorr. This was followed by the growth of an ultrathin BTO layer ($\sim$ 10 nm) at 700 $^0C$ under a higher oxygen pressure of 100 mTorr. After deposition, the samples were slowly cooled to room temperature over $\sim$ 4 h in an oxygen atmosphere of 100 mTorr to preserve oxygen stoichiometry and structural quality. The as-grown BTO/LSMO films were then transferred ex situ to the sputtering chamber for the subsequent deposition of the Py top layers.  Subsequently, the Py top layers were deposited from a Permalloy (Py) target to fabricate Py/BTO/LSMO heterostructures on MgO (100) and LAO (100) substrates. The sputtering process was carried out under a base pressure of $\sim$ 1 $\times$ 10$^{-7}$ mbar, with an Ar working pressure of $\sim$ 0.05 mbar and an applied power of 35 W. During deposition, the substrates were continuously rotated to ensure uniform film thickness and were maintained at 450 $^0C$ to promote high-quality crystalline growth. Following deposition, the samples were annealed in situ at 450 °C for 1 h under high vacuum ($\sim$ 1 $\times$ 10$^{-7}$ mbar), and subsequently cool down gradually to room temperature. The film thickness was monitored in real time using a quartz crystal microbalance. The thickness of the Py layer was fixed at $\sim$ 14 nm, corresponding to a deposition rate of $\sim$ 0.6$\SI{1}{\angstrom}$.\\
The structural quality and epitaxial growth of the films were examined by reflection high-energy electron diffraction (RHEED; STAIB Instruments, operated at 30 keV and analyzed using KSA400 software) and X-ray diffraction (XRD; Rigaku SmartLab employing Cu K$_\alpha$ radiation with $\lambda$ = 1.5406 $\SI{1}{\angstrom}$, equipped with a five-axis goniometer). The structural imaging was also done by using High-Angle Annular Dark-Field Scanning Transmission Electron Microscopy (HAADF-STEM) on aberration corrected Thermofisher TITAN Themis 300 at 300 kV with 24.5 mrad convergence angle at 160 mm camera length. Magnetic and magnetotransport properties, including M–H hysteresis loops and Hall resistivity, were measured using a Physical Property Measurement System (PPMS, Quantum Design DynaCool 9 T). The diamagnetic background arising from the substrates was carefully subtracted from all magnetization measurements.\\
The ferroelectric polarization behavior was further examined using piezoresponse force microscopy (PFM) performed on an MFP-3D Origin system equipped with a Ti/Ir-coated conductive probe (Asylum Research, ASYELEC.01-R2) having an approximate tip radius of 2.8 nm, spring constant of $\sim$2.8 N m$^{-1}$, and resonance frequency near 75 kHz. The acquired PFM images were subsequently analyzed using the WSxM software package.

\subsection{Computational methods}
The theoretical investigation of the ferroelectric BaTiO$_3$ is performed using first-principles calculations within density functional theory (DFT) with the projector augmented wave (PAW) method \cite{paw}, as implemented in VASP \cite{vasp}. The exchange-correlation functional within the generalized gradient approximation (GGA) \cite{gga,gga2} has been used for the calculations. Energy cutoff of 800 eV and a regular 8$\times$8$\times$8 Monkhorst-Pack \cite{m.pack} k-point mesh for the Brillouin zone integrations are used to achieve electronic convergence.\\
We have used lattice parameters obtained from the recent experiments \cite{Yadav2025} with the atomic positions taken from an earlier work \cite{Acta.Cryst} as a starting point. The lattice parameters obtained from the experiments are a = 3.95 \AA\ and c = 4.06 \AA. We further relax the atomic positions, keeping the cell shape and cell volume fixed until the Hellmann-Feynman forces on all atoms are reduced below  0.001 eV\AA$^{-1}$. Using the relaxed structure, we extract the atomic displacements from their inversion-symmetric positions, which we refer to as $\delta$. Following this, we have artificially constructed several structures by changing the $\delta$ values to understand the effect of the atomic displacements, $\delta$, on the electronic structure of BaTiO$_3$. The band structure and spin projections for these structures are then calculated in the presence of spin-orbit interactions.\\
The atomic-site electric dipole moment within an atomic sphere around each atom in BaTiO$_3$ is computed by decomposing the density matrix $\rho_{{l_1}{m_1}{l_2}{m_2}}$, obtained from a self-consistent calculation, into irreducible spherical tensor components $w^{kpr}_t$ \cite{multipole,multipole2}. Here $k$, $p$, $r$, and $t$ represent the spatial index, spin index, rank of the tensor, and the tensor component, respectively. The spatial index denotes the spatial part of the multipole, so for a dipole, $k=1$. The spin index $p$ can be either 0 or 1, depending on the presence and absence of time-reversal symmetry. Hence, we have $p=0$ for an electric dipole moment. The rank $r$ can vary from $|k-p|$ to $k+p$, so for an electric dipole, $r=1$, indicating that it is a vector quantity. For a given $r$, there are $(2r+1)$ components, given by $t=-r,...r$. For the electric dipole moment, this implies $t= -1, 0, {\rm and}\ 1$, leading to the spherical tensor components, $w^{101}_{-1}$, $w^{101}_{0}$, and $w^{101}_{1}$, which represent the $y$, $z$, and $x$ components of the dipole moment, respectively. The electric dipole moment breaks the inversion symmetry, and, therefore, only those density matrix elements contribute for which $l_1+l_2$ corresponds to an odd number.

\section*{acknowledgements}
D.M. and S.B. thank the Anusandhan National Research Foundation, Government of India (Grant No. ARNF/ARG/2025/007161/PS). A. R., M.G.,  and S.B. thank National Supercomputing Mission for providing computing resources of ‘PARAM Rudra’ at IIT Bombay, implemented by CDAC and supported by the Ministry of Electronics and Information Technology (MeitY) and Department of Science and Technology, Government of India. S.B. gratefully acknowledges financial support from the IRCC Seed Grant (Project Code: RD/0523-IRCCSH0-018), the INSPIRE Research Grant (Grant No.- DST/INSPIRE/IFF/BATCH-20/2024-25/IFA 23-PH 299), and the ANRF PMECRG Grant (Grant No.- ANRF/ECRG/2024/001433/PMS). D.M.  acknowledges financial support from the India–Russia Joint Research Call, DST, Government of India (Grant No. DST/IC/RSF/2024/542).
\textbf{Preprint Notice: This manuscript has been submitted to Advanced Functional Materials and is currently under consideration for publication. The present version is a preprint that has not been peer reviewed. Subsequent versions of this work may differ from the final published article.}
\section*{Data Availability Statement}
The data that support the findings of this study are available from the corresponding author upon reasonable request.

\section*{Conflict of Interest Disclosure}
The authors declare that they have no known competing financial interests or personal relationships that could have appeared to influence the work reported in this paper.

\section*{Permission to Reproduce Material from Other Sources}
No material from other sources requiring permission was reproduced in this work.

\section*{Supporting Information}


\printendnotes

\bibliography{main}

\href{run:Supporting_Information.pdf}{Supporting Information}


\end{document}